\documentclass[aps,prb,groupedaddress,twocolumn]{revtex4}
\usepackage{graphicx}
\usepackage{multirow}
\usepackage{epstopdf}
\begin{document}
\title{ Mass transport of adsorbates near a discontinuous structural phase transition
}

\author{E. Granato $^{1,2}$ , S.C. Ying $^2$,  K.R. Elder $^{3,4}$, T. Ala-Nissila $^{2,4}$ }

\address{$^1$Laborat\'orio Associado de Sensores e Materiais,
Instituto Nacional de Pesquisas Espaciais,12227-010 S\~ao Jos\'e dos
Campos, S\~ao Paulo, Brazil}
\address{$^2$Department of Physics, P.O. Box 1843, Brown University,
Providence, RI 02912-1843, USA}
\address{$^3$Department of Physics, Oakland University, Rochester,
Michigan 48309-4487, USA}
\address{$^4$COMP CoE at the Department of Applied Physics, Aalto University School of Science, P.O. Box 11000,
FI-00076 Aalto, Espoo, Finland}

\begin{abstract}
We study the mass transport dynamics of an adsorbed layer near a discontinuous
incommensurate striped-honeycomb phase transition via numerical simulations of 
a coarse-grained model focusing on the motion of domain walls rather than
individual atoms.  Following an initial step profile created in the
incommensurate striped phase, an intermediate hexagonal incommensurate phase
nucleates and grows, leading to a bifurcation into two sharp profiles propagating in 
opposite directions as opposed to broad profiles  induced by atomic diffusive
motion.  Our results are in agreement with recent numerical simulations of a
microscopic model as well as experimental observations for the Pb/Si(111)
adsorbate system. 
\end{abstract}


\maketitle

\section{Introduction}

Recently, there have been extensive
studies of both the statics and dynamics of the Pb/Si(111) 
system \cite{Man,conf,Man2004}. At equilibrium, the system can exist in a striped
incommensurate (SI) phase with stripes of domain walls separating commensurate domains, 
as well as in a hexagonal incommensurate (HI)
phase with a hexagonal pattern of domain walls \cite{Man2004}. In growth processes, the system
displays spontaneous self-organization and height selection
of Pb islands beyond the monolayer regime. The most striking feature of this system is that the 
observed rate of island growth implies a rate of mass
transport orders of magnitude faster than that from
the usual atomic diffusion mechanism \cite{conf,Man2004,fasttransport}.  
Theoretical models \cite{Wang,gy12,grana13} indicate that domain wall motion in
an incommensurate phase can provide the basic mechanism for such fast
dynamics. This anomalous fast mass transport dynamics  was subsequently
confirmed in another experimental study  following the refilling of a hole region in the adsorbate layer in real time \cite{tringides13}. The results also showed an unexpected bifurcation of
 the initial step profile into two sharp fronts, with a hexagonal phase in between, propagating in
opposite directions at a speed much faster than that due to simple atomic diffusion. 

Previously, we performed a molecular dynamics (MD) simulation \cite{grana13}
study of an atomistic model that admits both the SI and HI phases. We found
that for an initial step profile separating a bare substrate region (or a 
hole) from the rest of the SI
phase, the  domain wall dynamics leads to a bifurcation of the initial step
profile into two interfaces  propagating in
opposite directions at a superfast speed with a HI phase in between, in agreement with the experimental observation on the Pb/Si(111) system
\cite{tringides13}. This
theoretical study  indicates that  there are two central ingredients for the observed anomalous superfast mass transport mechanism
with profile bifurcation. The first is the existence of a discontinuous transition between two
incommensurate phases such as the SI  and the HI phases corresponding to different coverages. 
The second is  the ability of the SI phase to transform itself rapidly into the HI phase near the boundary of the two phases, and the
ultrafast domain wall dynamics in the HI phase with a negligible
Peierls pinning barrier. However, the simulation study was limited to
relatively small system sizes and short time scales when compared to the experimental systems, and the propagating  fronts observed in the simulation studies was not as sharp as the experimentally observed. 
To  overcome the system size and time scale limits and clarify the basic physics
behind the observed  anomalous mass transport mechanism  like the one observed for Pb/Si(111) system, we consider in this work a simple
continuous density field description of a strained overlayer by the Phase Field
Crystal (PFC) model \cite{elder04}. Unlike the conventional PFC model, which
retains density variation at microscopic atomic length scales, here we employ a
coarse-grained PFC model where the fundamental length scale corresponds to the
separation between the domain walls. Thus the origin of the formation of
domains and domain walls due to the competition of lattice mismatch strain
energy and the adsorbate-substrate binding energy do not appear explicitly in
the model.  Instead, a periodic array of domains in the incommensurate phase is
built into the model via a preferred length scale that corresponds to the
separation between the domain walls. This model allows for both an SI and a
honeycomb incommensurate (HoI) phases. There is a discontinuous transition
between the SI and the HoI phases. This will lead to the bifurcation of the
propagating fronts just as  that observed in the Pb/Si(111) system
\cite{tringides13} resulting from the  discontinuous SI-HI transition. The PFC
model also has negligible conversion barriers between the SI and HoI phases as
well as that for the Peierls barrier for the HoI model, which are the other
ingredients for the anomalous mass transport mechanism. The main advantage of
this  simple coarse-grained model is that it allows us to study much larger system
sizes and get a clear qualitative physical picture of the mass transport
mechanism in these systems.

\section{Coarse-grained Phase-field crystal model }

The long-time dynamics of the adsorbed overlayer in 
the incommensurate phase is essentially controlled by the nature and
interaction of the topological defects that characterize such a phase, which
consist of an array of interacting domain walls forming the SI, HI or HoI
phases.  To model such topological defects in the simplest way, we use  a
phase-field description, where
the physically relevant continuous density field is the adsorbed layer
coverage. Phase-field models are based on free-energy functionals, which are
constructed by considering symmetries
and conservation laws \cite{langer}. In order to take into account the
structural changes of the domain-wall structure of the adsorbed
layer, we follow the approach of the two-dimensional phase-field crystal model
\cite{elder04}, 
described by the free-energy functional
\begin{eqnarray}
F && =  \int dx dy \{  \frac{1}{2} r (\rho -\rho_o)^2 + \frac{1}{2}(\rho -\rho_o ) (\nabla^2 + q_0^2)^2 (\rho-\rho_o) \cr
 && + \frac{1}{4} (\rho -\rho_o)^4 - \rho \  V( x,y ) \} ,
\label{model}
\end{eqnarray}
where $\rho(x,y)$ is the density field,  $r <0 $ and $q_o$  are effective dimensionless parameters, 
and $\rho_o$ is a dimensionless reference density. 
For convenience, we set  $\rho_o=1$ and $q_o=1$. The fundamental length scale
is set by ${2\pi}/{q_0}$ which corresponds to the spacing between domain walls.
The last term represents a pinning potential $V(x,y)$. 
Unlike the conventional PFC model \cite{elder04}, where the density field corresponds 
to the atomic density coarse-grained over vibrational time scales, we consider the 
present model as described in Eq. (\ref{model})  as a
coarse-grained description of the overlayer, which averages out spatial
variations at the microscopic  scales, but incorporates the domain wall
patterns. The domain walls are light (heavy)  for a compressively (tensile)
strained adsorbate layer.  
Correspondingly, for a compressively strained system the regions near maxima in
the phase field $\rho(x,y)$ correspond to a commensurate domain, whereas the
region around the minima of the density constitute the domain walls.  For a
tensile strained overlayer, we just need to reverse the interpretation of the
maxima and minima of the density as domain walls and commensurate domains,
respectively.  Note that in this interpretation of the model,  there is no
atomic spatial resolution, but only the spatial resolution of the domain wall
structure. It does incorporate the essential ingredient for fast mass
transport with profile bifurcation, which  is the existence of a structural phase transition with
discontinuity in the density.
Just as in the standard PFC model \cite{elder04}, the model of  Eq.
(\ref{model}) displays a first-order transition between the SI and HoI
phases with light (heavy) domain walls for decreasing (increasing) density.  


The main assumption for the dynamics  is that the density field $\rho(x,y,t) $ should evolve in time in a way that reduces the total free
energy $F$. Since  density field is conserved, it satisfies the continuity equation
\begin{equation}
\frac{\delta \rho}{\partial t} = - \nabla \cdot \vec J ,
\label{cont}
\end{equation}
where the current density is given phenomenologically by
\begin{equation}
\vec J = - \Gamma \nabla \frac{\partial  F }{\partial \rho},
\label{fick}
\end{equation}
where $\Gamma$ is a kinetic coefficient setting the fundamental time scale for
the domain wall motion. This should be orders of magnitude smaller than the
atomic diffusion time scale at low temperatures since it is controlled by the
relatively small Peierls energy barrier \cite{pokrov} pinning the domain walls and governing the conversion of the SI phase to the HoI phase, 
rather than the corrugation of the adsorption potential which controls atomic
diffusion.  Due to the discontinuous transition described by Eq. (\ref{model}),
the dependence of the current density $\vec J$ on the density field $\rho$ does
not follow, in general,  the usual Fick' s law
$\vec J = - D \nabla \rho$. This is consistent with the behavior found in the
experiments for Pb/Si(111), which has been argued  \cite{tringides13} to imply
an apparent anomalous diffusion. From Eqs. (\ref{cont}) and (\ref{fick}), the
time evolution of $\rho$ is then described by the Cahn-Hilliard dynamic
equation \cite{cahn}
\begin{equation}
\frac{\partial \rho}{\partial t} =\Gamma  \nabla^2 \frac{\partial  F }{\partial \rho}.
\label{dynamics}
\end{equation}

\section{Numerical results}

\begin{figure}
\centering
\includegraphics[width=0.9\columnwidth]{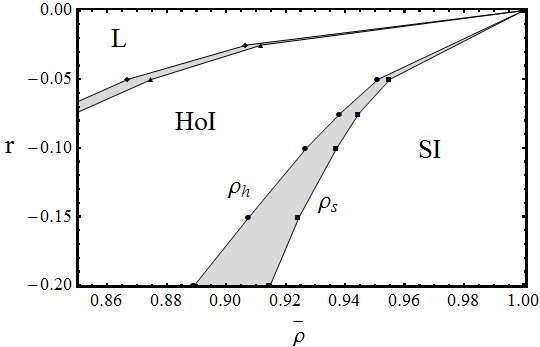}
\caption{Phase diagram showing the striped incommensurate phase (SI),
honeycomb incommensurate phase (HoI),  and the coexistence region (dark area)
in the range $\rho_h < \bar \rho < \rho_s$. L corresponds to a uniform phase
without domain-wall patterns.}
\label{phd}
\end{figure}

\begin{figure}
\centering
\includegraphics[width=\columnwidth]{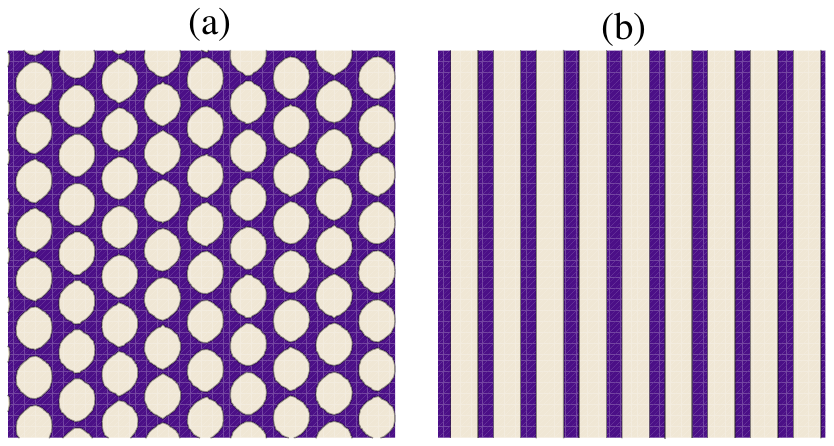}
\caption{Domain wall patterns corresponding to the (a) honeycomb incommensurate
phase and (b) striped incommensurate phase.  The dark areas correspond to
domain wall regions where the phase field $\rho(x,y)$ is closest to its minimum
value. }
\label{phases}
\end{figure}

The time evolution  was determined by  numerical integration of  the dynamical
equation, Eq. (\ref{dynamics}),  on a uniform square grid of size $ L_x dx
\times L_y dy $ with $d x =d y = \pi/4$ and $L_x=L_y=256 -512$,  and time steps
$d t= 0.05 - 0.1$.  
Figure \ref{phd} shows a portion of the  phase diagram near the SI to HoI phase
transition as a function of the average density $\bar \rho$ (for $V(x,y)=0$)
for the light domain wall case. In the range $\rho_h <\bar \rho <  \rho_s$, the
honeycomb and striped phases coexist while for $\bar \rho < \rho_h$ and $\bar
\rho  > \rho_s$, the equilibrium phases correspond to the HoI and SI phases,
respectively, as shown in Fig. \ref{phases}.  For small $|r| $ or small $\bar
\rho$, there is also a uniform phase without domain-wall patterns, which is of
no interest here.  The time evolution of an initial state with a density
profile containing a hole with a lower density
 ($\rho < \rho_h$) will be different for an initial SI or a HoI phase.  For an
initial SI phase, the decrease in the average density after creating the hole
can bring the system  near or into the coexistence region.  If the average
density $\bar \rho$ after the creation of the hole is in the range  $ \rho_h <
\bar \rho < \rho_s$, then an HoI region centered at the hole can coexist with
the remaining SI phase at long times. For an initial HoI phase, however, the
decrease in the density moves the system further away from the coexistence
region and there is just a spreading of the density without an expanding interface, 
following the creation of a hole, eventually tending to a uniform profile.

\begin{figure}
\centering
\includegraphics[width=\columnwidth]{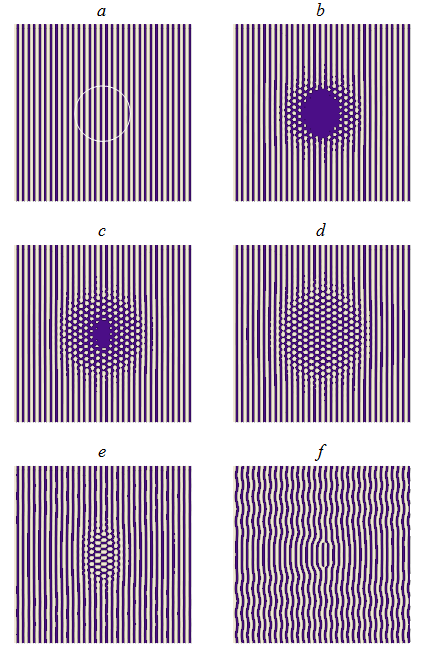}
\caption{ Snapshots of the density field when a "hole", namely,  a circular
region of radius $R=40 dx $ (white circle) with an average density $\bar \rho
=0.6< \rho_h $, is introduced in the stripe
phase, for increasing times:  (a)  $t=0$, (b) $t=6$, (c) $t=9$, (d) $t=21$ ,
(e) $t=78$, and (f) $t=126$ in units of $4.8 \times 10^5 dt $.  The initial
striped phase has an
average density of $\bar \rho = 0.98$  and $ \bar \rho=0.951$ after introduction of the hole.}
\label{growtha}
\end{figure}

\begin{figure}
\centering
\includegraphics[width=\columnwidth]{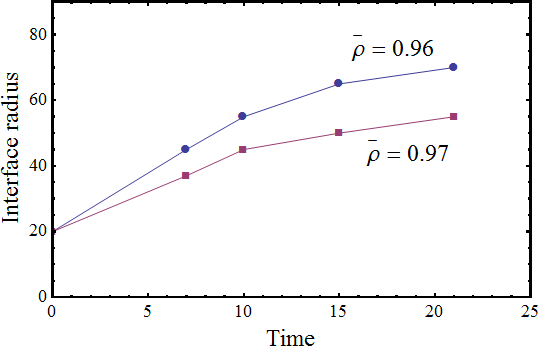}
\caption{ Radius  of the expanding SI-HoI interface as a function of time (in
units of $ 4.8 \times 10^5 dt $) for initial stripe phases with  different
average densities $\bar \rho $, for a hole of radius $R=20 dx$.  }
\label{expand}
\end{figure}

We will consider in detail an initial state in the SI phase when the density is
higher but close to the coexistence phases boundary $\rho_s$. From now on we
set the parameter $r=-0.1$. 
In Fig. \ref{growtha}  we show snapshots of the density field for increasing times when  a hole 
with local density $\rho < \rho_h$ is created in an initial striped phase such
that the average density $\bar \rho$ after the creation of the hole is still
higher than $\rho_s$.  An intermediate HoI phase starts to nucleate around the
edge of the hole  and grows with time, leading to a bifurcation of the initial
step profile at edge into two
profiles propagating in opposite directions. The outward front  corresponds  to
a SI-HoI interface, where the local stripe pattern is converted into  a
honeycomb pattern, while the inwards front is a step edge refilling the hole.
However, the resulting HoI region decays  back into a SI phase for sufficiently
long times (Figs. \ref{growtha}e  and \ref{growtha}f). As shown in Fig.
\ref{expand}, the time evolution of the radius of the expanding circular
interface depends on the density of the initial striped phase, being faster for
an initial density closer to the SI-HoI phase boundary, $\rho_s$, of the
coexistence region in the phase diagram of Fig. \ref{phd}.  

\begin{figure}
\centering
\includegraphics[width=\columnwidth]{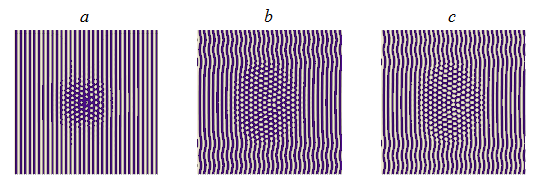}
\caption{ Snapshots of the density field for increasing times when the average
density after introducing a hole with density $\bar \rho=0.8 $ is inside the
coexistence region. The average density of the initial striped phase is $\bar
\rho = 0.945$ , and $\bar \rho=0.934$ after the introduction of the hole. (a)
$t=4$, (b) $t=21$, and  (c) $t=126$  in units of $4.8 \times 10^5 d t$.  
  }
\label{growthb}
\end{figure}

For comparison, in Fig. \ref{growthb} we show the time evolution, in the same
time interval, when the average density after the introduction of the hole is
within the coexistence range,  $ \rho_h < \bar \rho < \rho_s$.  Here the HoI
region centered at the hole remains at long times. 

The nucleation and growth of the intermediate HoI phase in Fig. \ref{growtha} and
the time evolution of the radius of the expanding HoI-SI interface in Fig.  \ref{expand}
are qualitatively consistent  with MD simulations of an atomistic model \cite{grana13} and with 
experimental observations for the  Pb/Si(111) system \cite{tringides13}.  A sign of the decay
of the  HoI region at long times could be the partial recovery of density in the hexagonal phase
found in the  experiment.  For this system, the boundary of the coexistence region between HoI phase 
and SI phase, at coverage values $\rho_h$ and $\rho_s$ in Fig. \ref{phd}, should correspond  
to the experimentally observed discontinuous jump at the hexagonal-stripe phase boundary between
coverage $\theta  \approx 1.26 $  ML and $\theta \approx 1.28 $ ML .  In this system, fast dynamics 
are observed experimentally at lower coverages,
as long as it exceeds some critical value, $\theta_c \approx 1.24$  ML.  
The existence of this lower critical coverage for this system is most likely due to the existence of other 
commensurate phases at or slightly above coverage \cite{Man}  1.2 ML.  This is
beyond the scope of our simple model, which focuses only on the SI and HoI
phases near a single commensurate phase at a slightly higher coverage than the
SI/HoI boundary.  
Other more microscopic models \cite{gy12,Wang} can 
account for this critical coverage as a competition between the
lattice mismatch strain energy and the adsorbate-substrate
binding energy for increasing coverage. 

\begin{figure}
\centering
\includegraphics[width=\columnwidth]{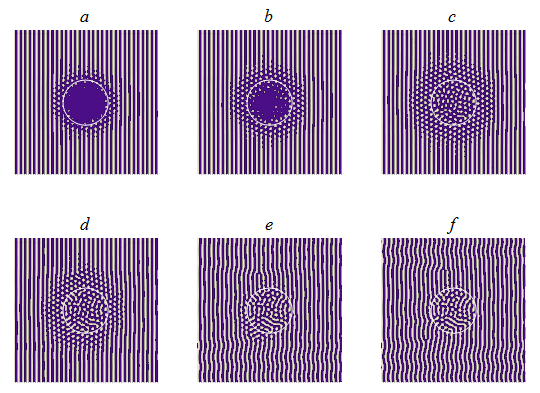} \caption{ Snapshots of the
density field with quenched disorder inside the hole for increasing times: 
(a)  $t=5$, (b) $t=8$,  (c) $t=21$, (d) $t=42$, (e) $t=72$, and (c) $t=126$ in units of $4.8 \times 10^5 dt$.
Average density of the initial striped phase is $\bar \rho = 0.98 $,  and $\bar
\rho=0.943$ after the introduction of the hole
with density $\bar \rho=0.5 $. The disorder strength here is $ \Delta = 0.08$.}
\label{disor}
\end{figure}

In the refilling experiment for Pb/Si(111) \cite{tringides13},  the hole is not
empty. There is an initial density corresponding to a tightly bound layer of
low coverage (1/3 monolayer $\beta$ phase) in the hole, which is only partially
equilibrated. To mimic the effect of this partially equilibrated layer, we
allow for a random, quenched
pinning potential $V(x,y)$ in Eq. (\ref{model}) localized only inside the hole.  
We take the simplest model for the random potential, defined 
by the correlations
\begin{equation}
\langle V(x,y) V(x',y') \rangle = \Delta^2 \delta(x-x') \delta (y-y') ,
\end{equation}
where $\Delta$ is  a measure of the strength of the disorder.  As shown in
Fig. \ref{disor}, disorder inside the hole leads to a distorted honeycomb phase
inside the hole with 
structural defects. For holes of sufficiently  larger sizes, this should
correspond to an amorphous glassy phase even for weak disorder strength
\cite{grana2011}. Such a phase corresponds to the disordered phase around the
inner refilling edge  observed  experimentally \cite{tringides13}.  This leads
to a static hexagonal-amorphous interface between the two profiles propagating
in opposite directions.  However, the hexagonal intermediate phase decays back into
the SI phase for sufficiently long times (Fig. \ref{disor}f).  Interesting enough,  the
amorphous phase inside the hole still remains at such long times. 

\section{Summary and Conclusions}

In this work, we have presented numerical results based on an appropriately
coarse-grained PFC model to illustrate the basic physics behind bifurcation of
the initial coverage profile in the fast mass transport mechanism  observed
experimentally \cite{Man,conf,tringides13,Man2004}  for the Pb/Si(111)
adsorption system. 
The new model is similar to the traditional PFC model \cite{elder04}, but the
interpretation of the phase field and the fundamental length scale are
different. It focuses on the domain wall pattern and not the density variation
inside the commensurate domains at a microscopic scale. It shares with the
conventional PFC model the advantage that numerical work can be performed for
system sizes orders of magnitude larger than in microscopic MD simulation
studies \cite{grana13}. 
Our results for the mass transport mechanism are qualitatively similar to the previous MD
work of an atomistic model 
\cite{grana13}. Taken together, they clearly demonstrate that the  essential
ingredient for the mass transport with a bifurcation of the initial
profile is the presence of two incommensurate phases with a first order transition between the 
two incommensurate phases involving  a discontinuity in the coverage. In the work
presented here, the two incommensurate phases involved are the SI phase and the HoI phase, but
qualitatively it has the same feature as the SI-HI phase transition in the 
Pb/Si(111) adsorption system. This mass transport mechanism is fast because  the HI and the HoI phases have negligible Peierls pinning barriers while the conversion of the SI phase to the HI or HoI phase near the phase transition boundary also involves barriers much lower than those for atomic diffusion. The SI-HI transition corresponds to the
Pb/Si(111) adsorption system \cite{Man2004} and many heteroepitaxial  metallic
overlayers \cite{elder12}, while the SI-HoI transition occurs for a system such
as Xe/Pt(111), Xe/Graphite and Kr/Graphite \cite{BruchRev}. 
In these cases, the commensurate state is a  
$(\sqrt{3}\times\sqrt{3})R30^{\circ}$ phase, which can undergo a transition
into the SI phase and then to the HoI phase \cite{coper,elder16}.  Our results
here demonstrate  that the phenomena of fast mass transport should not be just
confined to the Pb/Si(111) system alone, but is expected to be a general
feature for a wide class of surface adsorption systems under appropriate
conditions.

\medskip

\acknowledgements
This work was supported by   S\~ao Paulo Research Foundation (FAPESP, Grant No.
2014/15372-3) (E.G), CNPq (E.G.), the Watson Institute at 
Brown University under a  Brazil Collaborative Grant (S.-C.Y.) and the National 
Science Foundation under Grant No. DMR-1506634 (K.R.E.). T.A-N. has been supported in part
by the Academy of Finland through its COMP Centre of Excellence Program 
(projects No. 251748 and No. 284621).


 \end{document}